\documentstyle[amsfonts,12pt,thmsa,sw20lart]{article}

\input{tcilatex}
\begin{document}

\author{A. Robledo \\
Instituto de F\'{i}sica, Universidad Nacional Aut\'{o}noma de M\'{e}xico,\\
Apartado Postal 20-364, M\'{e}xico 01000 D.F., Mexico.}
\title{Criticality in non-linear one-dimensional maps: RG universal map and
non-extensive entropy }
\date{}
\maketitle

\begin{abstract}
We consider the period-doubling and intermittency transitions in iterated
nonlinear one-dimensional maps to corroborate unambiguously the validity of
Tsallis' non-extensive statistics at these critical points. We study the map 
$x_{n+1}=x_{n}+u\left| x_{n}\right| ^{z}$, $z>1$, as it describes
generically the neighborhood of all of these transitions. The exact
renormalization group (RG) fixed-point map and perturbation static
expressions match the corresponding expressions for the dynamics of
iterates. The time evolution is universal in the RG sense and the
non-extensive entropy $S_{Q}$ associated to the fixed-point map is maximum
with respect to that of the other maps in its basin of attraction. The
degree of non-extensivity - the index $Q$ in $S_{Q}$ - and the degree of
nonlinearity $z$ are equivalent and the generalized Lyapunov exponent $%
\lambda _{q}$, $q=2-Q^{-1}$, is the leading map expansion coefficient $u$.
The corresponding deterministic diffusion problem is similarly interpreted.
We discuss our results.
\end{abstract}

\section{Introduction}

The proposal, in 1988, by Tsallis \cite{tsallis0} of a non-extensive
generalization of statistical mechanics unleashed an unprecedented
discussion \cite{tsallis1} on the foundations of this most basic branch of
physics. This is an exceptional event, since its inception, with Boltzmann
and Gibbs (BG), in the long and heretofore unchallenged history of
statistical mechanics. Many recent studies \cite{tsallis1} have offered
experimental and numerical evidences indicating limiting conditions for the
domain of validity of the canonical BG theory. But, so far there have been
no firm developments to tilt the balance towards universal acceptance of
these new ideas, nor a definite understanding of the deep-lying physical
reasons (believed in some cases to be a breakdown in the chain of increasing
randomness from non-ergodicity to completely developed chaoticity) for the
failure of the BG statistics and the competence of the non-extensive
generalization. However, here we present exact analytical results for a
class of critical points in non-linear maps from which a rigorous
corroboration and understanding of the non-extensive theory can be drawn 
\cite{latora1} - \cite{baldovin3}.

The renowned period-doubling and intermittency routes to chaos are now
classical examples of how the use as starting points of simple non-linear
discrete maps have often led to important developments in the theory of
non-linear dynamical systems \cite{schuster1}. The discovery of the
universal properties associated to these mechanisms, comparable with those
of conventional critical phenomena in statistical physics, triggered, over
two decades ago, an explosion of activity in the field and now both routes
are well understood and verified experimentally. Here we might add yet other 
{\it universal} aspects to their properties (explained in more detail below)
the non-extensivity marked by the power-law growth or decay of the
sensitivity to initial conditions, and the entropy extrema associated to
each of the fixed points that define the RG universality classes.

The well-known period-doubling and intermittency transitions are based on
the pitchfork and the tangent bifurcations. In particular, the
one-dimensional logistic map and its generalization to non-linearity of
order ${\frak z}>1$, $f_{\mu }(x)=1-\mu \left| x\right| ^{{\frak z}}$,$%
\;-1\leq x\leq 1$, exhibit these two different types of bifurcations an
infinite number of times as the control parameter $\mu $ reaches critical
values. At these critical points universal scaling laws and indexes hold,
independently of the details of the map, and are therefore shared by all
maps of the same order ${\frak z}$. As with other statistical mechanical
systems an explanation of universality is provided via the renormalization
group (RG) method, which has been successfully applied to this type of
iterated map. The RG doubling transformation, consisting of functional
composition and rescaling, was first devised to study the cascade of
period-doubling transitions and its accumulation point, and was later
applied to the intermittency transition. As pointed out \cite{schuster1} the
latter case is one of the rare examples where the RG equations can be solved
exactly. Here we indicate that this analytic solution holds generally not
only for tangent bifurcations (as known time ago \cite{schuster1}) {\it but
also} for pitchfork bifurcations (as noticed recently \cite{robledo0}). We
also point out that the fixed-point map can be determined as an extremum of
the non-extensive entropy, thus providing evidence for a possible connection
between extremal properties of entropy expressions and those for the fixed
points obtained from the usual RG recursive method \cite{robledo1}.

As required, at each of the map critical points the Lyapunov $\lambda _{1}$
exponent vanishes, and the sensitivity to initial conditions $\xi _{t}$ for
large iteration time $t$ ceases to obey exponential behavior, exhibiting
instead power-law behavior \cite{oldrefs1}. As the means to describe the
dynamics at such critical points, and based on the non-extensive entropy of
Tsallis $S_{Q}$ \cite{tsallis1} (see Eq. (\ref{tsallis1}) below), the $q$%
-exponential expression 
\begin{equation}
\xi _{t}=\exp _{q}(\lambda _{q}t)\equiv [1-(q-1)\lambda _{q}t]^{-1/(q-1)}
\label{sens1}
\end{equation}
containing a $q$-generalized Lyapunov exponent $\lambda _{q}$ was proposed 
\cite{tsallis2}, together with generalizations for the rate of entropy
production (referred here as the Kolmogorov-Sinai (KS) entropy \cite
{schuster1}) $K_{Q}$ (see Eq. (\ref{KSq1}) below) and for the Pesin identity 
$\lambda _{q}=$ $K_{Q}$ \cite{tsallis2}. The standard expressions are
recovered when $q,Q\rightarrow 1$. The use of the two indexes $q$ and $Q$ in
our application will become clear below. Furthermore, recent studies \cite
{tsallis2}-\cite{lyra1}, focusing mainly at the onset of chaos (the
accumulation point of the period-doubling bifurcations) of the ${\frak z}$%
-logistic map have revealed a series of connections between the Tsallis
entropic index $q$ and the non-linearity ${\frak z}$, the fractal dimension $%
d_{f}$ of the chaotic attractor, and the end points of its multifractal
singularity spectrum $f(\alpha )$. Here we identify $q$ and $\lambda _{q}$
in terms of the tangent and pitchfork critical map parameters.

The properties of dynamical systems at critical points can also be probed by
considering the deterministic diffusive processes that take place in
non-linear maps with discrete translational symmetry. These maps consist of
a sequence of cells containing each elements of a basic nonlinear map. One
such construction \cite{geisel1} makes use of the tangent bifurcation, and,
interestingly, anomalous diffusional behavior has been observed close to the
intermittency transition \cite{geisel1}, \cite{grigolini1}. Continuous-time
random-walk theory concepts have been applied \cite{geisel1}, \cite
{grigolini1} and, characteristically, the associated waiting-time
distribution function $\psi (t)$ has been shown to have non-exponential
power-law decay. The validity of the non-extensive approach to criticality
has been recently checked by examination of the anomalous diffusion at the
intermittency transition \cite{grigolini1}, \cite{grigolini2}. The key
results in these studies center on the analytical form predicted for $\psi
(t)$ which bears a close relationship to the power-law Eq. (\ref{sens1}).
Here we point out that the analytical forms for this and related
distributions actually correspond to those obtained by the repeated
application of the RG transformation. As mentioned, if the fixed point
solution can also be determined via an entropy optimization method the
expressions for these distributions are indicative of the same entropy
extremum.

\section{Outline of results}

The main points in the following analysis are:

1) The connection between the RG fixed-point map expression for the iterate $%
x_{t}$ associated to trajectories $x_{t}$ with $x_{0}\rightarrow 0$ at $t=0$
and the expression in Eq. (\ref{sens1}) for the sensitivity of initial
conditions $\xi _{t}$. This leads to simple identifications of the
non-extensive parameters as $q=2-z^{-1}$ and $\lambda _{q}\sim a$. We
corroborate the earlier known results $q=3/2$ and $q=5/3$ for the tangent
and pitchfork bifurcations for the logistic map ${\frak z}=2$, respectively 
\cite{tsallis2}.

2) The connection between the RG fixed-point map recursion relation for the
iterate $x_{t}$ and the expression obtained from $S_{Q}$ for the temporal
evolution of the number of cells $W_{t}$ occupied by an ensemble of initial
conditions spread over a finite-length interval $[0,x_{0}]$. (See the
definition of $S_{Q}$ in Eq. (\ref{tsallis1}), and the explanatory comments
at the end of this section about the relationship $q=2-Q^{-1}$ between the
indexes $q$ and $Q$). This leads to the identifications $Q=z$ and $K_{q}\sim
a$, confirming $q=2-z^{-1}$ and giving support to the validity of the
generalized Pesin identity $\lambda _{q}=$ $K_{Q}$ at the critical
transitions. See Ref. \cite{baldovin3} for a recent proof of the validity of
this generalized identity at the edge of chaos of the logistic map ${\frak z}%
=2$.

3) The general property of the non-extensive entropy function which
decreases monotonically as the RG transformation flows away from the fixed
point where it attains its maximum value.

4) The maximum entropy property of the distribution $\Psi (\tau )$, $\Psi
(\tau )=\int_{0}^{\tau }dt\ \psi (t)$, of cell residence time intervals $%
[0,\tau ]$ in the critical diffusive processes.

For clarity of presentation it is convenient to advance the following
property of the $q$-exponential function. For this function the ordinary
derivative $d\exp (x)/dx=\exp (x)$ and inverse $[\exp (x)]^{-1}=\exp (-x)$
properties become, respectively, $d\exp _{q}(x)/dx=[\exp _{q}(x)]^{q}$ and $%
[\exp _{q}(x)]^{-1}=\exp _{2-q}(-x)$, and these combine into the identity 
\begin{equation}
\exp _{q}(x)\equiv \left[ \exp _{Q}(\frac{x}{Q})\right] ^{Q},
\label{identity}
\end{equation}
where $q=2-Q^{-1}$. Thus, in a given problem where the $q$-exponential
function applies there appear related values of the $q$-index, that become
the same (and the above identity becomes trivial) as $q\rightarrow 1$.

\section{Critical dynamics from fixed-point map}

So, we recall \cite{schuster1}, \cite{hu1} the solution of Hu and Rudnick to
the Feigenbaum RG recursion relation obtained for the case of the tangent
bifurcation. For the transition to periodicity of order $n$ consider the $n$%
-th composition $f^{(n)}$ of the original map $f$ in the neighborhood of one
of the $n$ points tangent to the line with unit slope and shift the origin
of coordinates to that point. Then, one obtains 
\begin{equation}
f^{(n)}(x)=x+u\left| x\right| ^{z}+o(\left| x\right| ^{z}),  \label{n-thf1}
\end{equation}
where $u>0$ is the leading expansion coefficient (the right-hand-side in Eq.
(\ref{n-thf1}) may depend on $n$ only through the omitted terms). The RG
fixed-point map $x^{\prime }=f^{*}(x)$ was found to be 
\begin{equation}
x^{\prime }=x\exp _{z}(ux^{z-1})=x[1-(z-1)ux^{z-1}]^{-1/(z-1)},
\label{fixed1}
\end{equation}
as it satisfies $f^{*}(f^{*}(x))=\alpha ^{-1}f^{*}(\alpha x)$ with $\alpha
=2^{1/(z-1)}$ and has a power-series expansion in $x$ that coincides with
Eq. (\ref{n-thf1}) in the two lowest-order terms. (We have used $%
x^{z-1}\equiv \left| x\right| ^{z-1}{\rm sgn}(x)$). Eq. (\ref{fixed1}) can
be rewritten as $x^{\prime -(z-1)}=x^{-(z-1)}-(z-1)u$, a translation of the
power law $y=$ $x^{-(z-1)}$, the property that led Hu and Rudnick in their
derivation. The effect of a perturbation of the form $x^{-p}$ transforms
this into 
\begin{equation}
x^{\prime -(z-1)}+\varepsilon ^{\prime }x^{\prime -p}=x^{-(z-1)}+\varepsilon
x^{-p}-(z-1)u,  \label{fixed3}
\end{equation}
or, to lowest order in $\varepsilon ^{\prime }=\varepsilon $, 
\[
x^{\prime }=x\exp _{z}(ux^{z-1})-\frac{\varepsilon x^{z-p}}{z-1}\left[ \exp
_{z}(ux^{z-1})\right] ^{z} 
\]
\begin{equation}
+\frac{\varepsilon x^{z-p}}{z-1}\left[ \exp _{z}(ux^{z-1})\right] ^{z-p}.
\label{fixed4}
\end{equation}
The recurrence relation in Eq. (\ref{fixed3}) remains invariant under
iteration and rescaling by $\alpha $ when $\varepsilon ^{\prime
}=\varepsilon $ is multiplied by the factor $\gamma =2^{(p-z+1)/(z-1)}$. As
it is known \cite{hirsch1}, \cite{hu1}, the leading relevant perturbation
about this critical point is that which corresponds to a shift from
tangency, for this case the lowest order term in $x$ in Eq. (\ref{fixed4})
must be of order one, and so we fix $p=z$ for the perturbation exponent (in
the unormalized eigenfunction).

Fittingly, it can be readily concluded that the RG recursion relation in the
neighborhood of the pitchfork bifurcations $\left.
df^{(2^{k-1})}(x)/dx\right| _{x=0}=$ $-1$ conforms with the above scheme and
that its solution is {\it also} applicable to these transitions provided the
sign of $u$ is changed for $x>0$. It is interesting to point out that at the
time the RG method was first applied to the tangent bifurcation it was
thought that the simple closed form obtained for the fixed point map, the $q$%
-exponential in Eq. (\ref{fixed1}), would most likely prove not to be
fruitful for the study of period doubling \cite{schuster1}, \cite{hu1}. Not
only is this solution applicable to each period doubling transition with the
minor modification of the indicated sign change, but also, as it has been
recently demonstrated \cite{baldovin2}, the time evolution at their
accumulation point, the edge of chaos, is expressable in terms of similar $q$%
-exponentials. Below we comment on the caveats that appear in relating $z$
above to the ${\frak z}$ in the initial map e.g. $f_{\mu }(x)=1-\mu \left|
x\right| ^{{\frak z}}$.

When the number of iterations (after an initial $x_{0}$) is large and the
continuous time $t$ approximation is taken ($x^{\prime }=x_{t}$, $x_{0}=x$
and $u=at$ in Eq. (\ref{fixed1})) the fixed-point map can be written as 
\begin{equation}
x_{t}=x_{0}[1-(z-1)ax_{0}^{z-1}t]^{-1/(z-1)},  \label{fixed5}
\end{equation}
with properties $dx_{t}/dt=ax_{t}^{z}\;$and$%
\;dx_{t}/dx_{0}=(x_{t}/x_{0})^{z} $. Thus, the fixed-point and perturbation
relations can be used to describe the time evolution of intervals containing
iterates, an initial interval $[0,x_{0}<x_{+}]$ of length $\Delta x_{0}$
evolves into $[0,x_{t}<x_{+}]$ with length $\Delta x_{t}$ where $x_{t}$ is
the image of $x_{0}$ at time $t$. Accordingly, Eq. (\ref{fixed3}) with ($p=z$%
) becomes 
\begin{equation}
\Delta x_{t}^{-(z-1)}+\varepsilon _{t}\Delta x_{t}^{-z}=\Delta
x_{0}^{-(z-1)}+\varepsilon _{0}\Delta x_{0}^{-z}-(z-1)at.  \label{fixed6}
\end{equation}
We should like to emphasize that the expression for time evolution Eq. (\ref
{fixed5}) has the same universal form as that for the fixed-point map Eq. (%
\ref{fixed1}). In the neighborhood of the points of tangency $f^{(n)}(0)$
the iterates follow monotonic paths with time shape set by the map shape
itself, i.e. long time {\it dynamics} follow the {\it static} solution of Hu
and Rudnick. It should also be kept in mind that Eq. (\ref{fixed6}) applies
only to small perturbations as this corresponds to the perturbation
recursion relation obtained to linear order around the RG fixed point. We
also comment that our use here of the continuous time approximation (valid
for large iteration time $t$) in deriving Eq. (\ref{fixed1}) is only for
expediency reasons and has no effect in the generality of our results as
this equation is equally obtained from the scaling property 
\begin{equation}
x_{m}=f^{*^{(m)}}(x)=m^{-1/z-1}f^{*}(m^{1/z-1}x),\ m=1,2...,
\label{trajectory1}
\end{equation}
that in turn is obtained from $f^{*}(f^{*}(x))=\alpha ^{-1}f^{*}(\alpha x)$
with $\alpha =2^{1/z-1}$ \cite{baldovin1}.

\section{Sensitivity to initial conditions}

We have reached a point at which a straightforward proof of the validity of
Tsallis' Eq. (\ref{sens1}) can be obtained. This is done by direct
evaluation of the sensitivity to initial conditions from its definition $\xi
_{t}\equiv \lim_{\Delta x_{0}\rightarrow 0}\left( \Delta x_{t}/\Delta
x_{0}\right) $. From Eq. (\ref{fixed5}) and $dx_{t}/dx_{0}=(x_{t}/x_{0})^{z}$
one obtains 
\begin{equation}
\xi _{t}(x_{0})=[1-(z-1)ax_{0}^{z-1}t]^{-z/(z-1)},  \label{sens2}
\end{equation}
and comparison with Eq. (\ref{sens1}) yields $q=2-z^{-1}$ and $\lambda
_{q}(x_{0})=zax_{0}^{z-1}$\cite{grigolini1}.

We can also confirm the values of $q$ obtained numerically by Tsallis et al 
\cite{tsallis2} for both the tangent and pitchfork bifurcations when ${\frak %
z}=2$ . The composition $f_{\mu }^{(n)}$ for the ${\frak z}$-logistic map $%
f_{\mu }(x)=1-\mu \left| x\right| ^{{\frak z}}$,$\;{\frak z}>1$, at the
tangent bifurcations is $f_{\mu }^{(n)}(x)=x+ux^{2}+o(x^{2})$,$\ u>0$, and
from $z=2$ one has $q=3/2$. For the pitchfork bifurcations one has instead $%
f_{\mu }^{(n)}(x)=x+ux^{3}+o(x^{3})$, because $d^{2}f_{\mu
}^{(2^{k})}/dx^{2}=0$ at these transitions and $u<0$ is now the coefficient
associated to $d^{3}f_{\mu }^{(2^{k})}/dx^{3}<0$. In this case we have $z=3$
in $q=2-z^{-1}$ and one obtains $q=5/3$. Notably, these specific results for
the index $q$ are valid for all ${\frak z}>1$ and therefore define the
existence of only two universality classes for unimodal maps, one for the
tangent and the other one for the pitchfork bifurcations \cite{baldovin1}.

Further, if one uses the fact \cite{gaspard1} that the invariant
distribution of $f^{(n)}$ in Eq. (\ref{n-thf1}) is of the form $\rho (x)\sim
x^{-(z-1)}$, the average $\overline{\lambda }_{q}$ of $\lambda _{q}(x_{0})$
over $x_{0}$ yields the constant $\overline{\lambda }_{q}\sim a$. In
relation to this we observe that this average is compatible with 
\begin{equation}
\overline{\lambda }_{q}=\int dx\ \rho (x)\ln _{q}\left| \frac{df^{*}(x)}{dx}%
\right| ,  \label{Lyapunovav}
\end{equation}
the $q$-extension of the customary \cite{schuster1} expression for $\lambda
_{1}$ as the average of $\ln \left| df^{*}(x)/dx\right| $ over $\rho (x)$.
By taking the $q$-logarithm (where $\ln _{q}y\equiv (y^{1-q}-1)/(1-q)$ is
the inverse of $\exp _{q}(y)$) of both sides of 
\begin{equation}
\xi _{t}(x_{0})=[1-(q-1)\lambda _{q}(x_{0})t]^{-1/(q-1)},  \label{sens3}
\end{equation}
and by recalling the definition of $\xi _{t}(x_{0})$ one obtains 
\begin{equation}
\lambda _{q}(x_{0})=t^{-1}\ln _{q}\left| \frac{df^{*(t)}(x_{0})}{dx_{0}}%
\right| ,  \label{Lyapunov}
\end{equation}
where $\ln _{q}\left| df^{*(t)}(x_{0})/dx_{0}\right| \sim t$ as $\lambda
_{q}(x_{0})$ does not depend on $t$. (Notice that the limit $t\rightarrow
\infty $ of Eq. (\ref{Lyapunov}) is the $q$-extension of the regular
definition of $\lambda _{1}(x_{0})$ \cite{schuster1}). Next, use of the
fixed-point relation $f^{*(t)}(x_{0})=t^{1/1-z}f^{*}(t^{1/z-1}x_{0})$ and $%
t=(x/x_{0})^{z-1}$ into (\ref{Lyapunov}) leads to $\lambda
_{q}(x_{0})=(x/x_{0})^{-(z-1)}\ln _{q}\left| df^{*}(x)/dx\right| $ and Eq. (%
\ref{Lyapunovav}).

\section{Rate of entropy change}

Even though our basic results have been derived, it is useful to extend our
analysis and reveal more details about the dynamics at these critical
points. Our following arguments focus on $x_{t}$ rather than $dx_{t}/dx_{0}$
and imply $Q=z$, thus confirming $q=2-z^{-1}$. They also indicate that the
rate of entropy change is $K_{Q}\sim a$, suggesting the validity of the $q$%
-generalization of the Pesin $\overline{\lambda }_{q}=$ $K_{Q}$ at these
critical states. Therefore, let's consider a cell partition ($N\rightarrow
\infty $ cells of length $l\rightarrow 0$) in the 'entire' phase space $%
[0,x_{+}]$. Consider also the time evolution of occupied cells $W_{t}$ by an
ensemble of iterates with an {\it extended} initial occupation $W_{0}$ that
spans $[0,x_{0}<x_{+}]$. We write the cell distribution of iterates $p_{i}$
as 
\begin{equation}
p_{i}=\frac{1+\delta _{i}}{W},\;\sum_{i}^{W}\delta _{i}=0,  \label{distr1}
\end{equation}
as this is a convenient form when considering normalized perturbations to
the uniform distribution. It is a general expression when no further
conditions are imposed on the $\delta _{i}$ but below we require small $%
\delta _{i}$. When Eq. (\ref{distr1}) is introduced into the KS $q$-entropy
difference 
\begin{equation}
K_{Q}t=S_{Q}(t)-S_{Q}(0),\;t\;\text{large,}  \label{KSq1}
\end{equation}
where 
\begin{equation}
S_{Q}=\frac{1-\sum_{i}^{W}p_{i}^{Q}}{Q-1}  \label{tsallis1}
\end{equation}
is the Tsallis entropy, we obtain 
\begin{equation}
W_{t}^{-Q}\sum_{i}^{W_{t}}(1+\delta
_{i,t})^{Q}=W_{0}^{-Q}\sum_{i}^{W_{0}}(1+\delta _{i,0})^{Q}-(Q-1)K_{Q}t.
\end{equation}
To lowest order in $\delta _{i}$ this reduces to 
\[
W_{t}^{-(Q-1)}+\frac{1}{2}Q(Q-1)\delta _{t}^{2}W_{t}^{-Q}=
\]
\begin{equation}
W_{0}^{-(Q-1)}+\frac{1}{2}Q(Q-1)\delta _{0}^{2}W_{0}^{-Q}-(Q-1)K_{Q}t,
\label{weight1}
\end{equation}
with $\delta ^{2}\equiv \sum_{i}^{W}$ $\delta _{i}^{2}$.

Remarkably, Eqs. (\ref{weight1}) and (\ref{fixed6}) can be seen to be
equivalent if $Q=z$ and if $W_{t,0}\sim $ $\Delta x_{t,0}$, $K_{Q}\sim a$
and $\frac{1}{2}Q(Q-1)\delta _{t,0}^{2}\sim \varepsilon _{t,0}$. The
entropic index $Q$ is simply given by the non-linearity $z$ while $K_{Q}$ is
the expansion coefficient $a$ (and through the proviso that $\overline{%
\lambda }_{q}=$ $K_{Q}$ this would imply $\overline{\lambda }_{q}\sim a$).
Our comparison between Eqs. (\ref{weight1}) and (\ref{fixed6}) is consistent
in that each is the lowest-order expression valid for small perturbations,
the first to the uniform cell distribution of iterates and the second to the
fixed-point map solution. We can draw the following conclusion from the
association of Eqs. (\ref{weight1}) and (\ref{fixed6}): The entropy $%
S_{Q}^{*}$ for the fixed-point map is maximum, as $S_{Q}^{*}-S_{Q}=Q\delta
^{2}/2W^{q}>0$, and the distribution becomes uniform, $p_{i}=W^{-1}$. The RG
flow is away from the fixed-point since $\gamma >1$ and the entropy
decreases as the RG transformation is applied. The proposed equivalence
between Eqs. (\ref{weight1}) and (\ref{fixed6}) is justified by making the
following connection with Eq. (\ref{sens1}). For a uniform distribution $%
W_{t,0}\sim $ $\Delta x_{t,0}$, as the ratio of the number of cells $W_{t}$
occupied at time $t$ to the distance $\Delta x_{t}$ is the length $l$, and
in the limit $N\rightarrow \infty $ and $l\rightarrow 0$, so that $\Delta
x_{0}\rightarrow 0$, with say $\Delta x_{0}/l=b$, one obtains $W_{t}=b\xi
_{t}$. This explicit link between the occupation numbers $W_{t}$ and $\xi
_{t}$ is in line with the all-important connection between the loss of
information embodied by the KS entropy and the sensitivity to initial
conditions in non-critical regimes \cite{schuster1}, and with the parallel
connection suggested by Tsallis et al \cite{tsallis2}, \cite{tsallis3} for
the analogous quantities at critical points.

It is worth the mention that the use of $q$-exponential (and the inverse $q$%
-logarithm) functions in the derivation of the properties of a given system
will in general introduce pairs of conjugate indexes such as $q=2-Q^{-1}$
above. As pointed out, taking derivatives and inverse operations lead to
expressions involving one or the other of these simply related values of the
entropic index (that become trivially the same $q=Q=1$ for the usual
exponential and logarithmic functions). Consequently, while some theoretical
features are expressed through $q$ some others appear expressed via $Q$, but
in all cases their occurrence and meaning is apparent.

\section{Deterministic diffusion}

We comment now on the deterministic diffusion process that takes place for
the variable $x_{t}$ in the map $x_{t+1}=f(x_{t})$, where $f(m+x)=m+f(x)$, $%
m $ integer and $f(-x)=-f(x)$. Geisel and Thomae \cite{geisel1} found that
when the unit-cell map $f(0\leq x\leq 1/2)$ is chosen to have the
(intermittency transition) form $x_{t+1}=x_{t}+ax_{t}^{z}$ within $%
[0,x_{+}\leq 1/2]$ anomalous diffusion, of the dispersive type, is generated
when $z>2$ since then the mean-square displacement $<\Delta x^{2}(t)>\sim
t^{1/(z-1)}$. This behavior follows \cite{geisel1} after the distribution of
cell residence times $\psi (t)$ is found to be 
\begin{equation}
\psi (t)=ax_{+}^{z-1}[1+(z-1)ax_{+}^{z-1}t]^{-z/(z-1)}.
\end{equation}
Grigolini and co-workers \cite{grigolini1} recognized the relationship
between the expressions for $\psi (t)$ and $\xi _{t}$. They also analyzed,
and verified, the validity of the non-extensive scheme involving the $q$%
-generalized KS entropy its equality with the $q$-generalized Lyapunov
exponent \cite{grigolini1}, \cite{grigolini2}. Here we can recap these
results as follows: The properties of the diffusion process driven by the
intermittency transition are universal since they stem from the RG
fixed-point map expression. Namely, $\psi (t)$ is, besides normalization,
the derivative of the inverse fixed-point map $dx_{0}/dx_{t}$ with $%
x_{t}=x_{+}$ and is therefore inversely proportional to $\xi _{t}=$ $%
dx_{t}/dx_{0}$. The distribution $\Psi (\tau )=\int_{0}^{\tau }dt\ \psi
(t)\sim <\Delta x^{2}(\tau )>$ is in turn proportional to the inverse of the
fixed-point map $x_{0}(x_{\tau })$. Also, 
\begin{equation}
\Psi (\tau )=A[1+(z-1)ax_{+}^{z-1}\tau ]^{-1/(z-1)},
\end{equation}
has the form of the distribution obtained by optimization of the
non-extensive entropy with index $Q=z$ for fixed $ax_{+}^{z-1}\tau $. The
related entropy index $q=2-Q^{-1}=2-z^{-1}$ applies to the residence time
distribution $\psi (t)$.

\section{Summary and discussion}

Thus, for the unimodal maps analyzed we have indicated that the exact {\it %
static} solution of the RG equations for the tangent and pitchfork
bifurcations also describe the {\it dynamics} of iterates. Further, these
fixed-point expressions are the same $q$-exponential expressions describing
the temporal evolution of ensembles of iterates prescribed by the
non-extensive formalism for non-linear maps at criticality. Therefore the RG
fixed-point map for the universality classes determined by $z$ can be
constructed via Tsallis entropic arguments. The equivalence rests on the
derivation of Eq. (\ref{sens1}) exclusively from RG procedures. The study of
the pitchfork and period-doubling transitions has been expanded via detailed
derivation of their $q$-generalized Lyapunov exponents $\overline{\lambda }%
_{q}$ and interpretation of the different types of sensitivity $\xi _{t}$ 
\cite{baldovin1}. Likewise, the properties of the intricate trajectories at
the edge of chaos of the logistic map ${\frak z}=2$ have been analytically
obtained leading too to the determination of $\overline{\lambda }_{q}$ and
interpretation of the dynamics at the strange attractor \cite{baldovin2}.
Further, the validity of the $q$-generalized Pesin identity at this
important state has been recently rigorously proved \cite{baldovin3}. More
generally, we addressed here the properties of the fixed-point map with the
purpose of finding additional evidence \cite{robledo1} for an existing
relationship between the variational properties of entropy expressions and
the RG approach when applied to systems with scale invariance properties. We
found that the non-extensive entropy function decreases monotonically as the
RG transformation flows away from the fixed point where it attains its
maximum value.

The ergodic hypothesis lies at the foundation of statistical mechanics
implying that trajectories in phase space cover uniformly the entire
pertinent regions. But is this hypothesis always correct? Already many years
ago the answer to this question has been probed by the study of simple
dynamical systems with only a few degrees of freedom \cite{schuster1}.
Besides their uncomplicated description these systems display extremely
convoluted motion in phase space that is neither regular nor simply ergodic,
and the mechanism by which ergodicity emerges in these and more complex
deterministic systems has been effectively explored by studying the
sensitivity to initial conditions and the associated Lyapunov exponents \cite
{schuster1}. The distinction between periodic and chaotic motion is
signaled, respectively, by the long-time exponential approach or departure
of trajectories with close initial positions. Here we have analyzed the
borderline critical states in one-dimensional non-linear maps at which the
exponential sensitivity law stops working and find that the universal
dynamical behavior at these states precisely follow the predictions of the
generalized non-extensive theory.

The currently developed non-extensive generalization \cite{tsallis0}, \cite
{tsallis1} of BG statistical mechanics entails a rare examination of the
domain of validity of this most fundamental base of physics. The suggested
physical circumstances for which BG statistics fails to be applicable are
thought to be associated to states that lack the full degree of chaotic
irregular dynamics that probes phase space thoroughly and that is necessary
for true equilibrium. Such anomalous states are signalled by the vanishing
of the largest Lyapunov exponent and exhibit non-ergodicity or unusual
power-law mixing. One particular case in point is that of critical states in
nonlinear unimodal maps. Here impeded or incomplete mixing in phase space
(the interval $-1\leq x\leq 1$) arises from the special 'tangency' shape of
the map at the pitchfork and tangent transitions that produces the monotonic
trajectories given by Eq. (\ref{fixed5}). This has the effect of confining
or expelling trajectories causing anomalous phase-space sampling, in
contrast to the thorough coverage in generic states with $\lambda _{1}>0$.
By construction the dynamics at the intermittency transitions, describe a
purely nonextensive regime, since the map studied here, Eq. (\ref{n-thf1}),
does not consider access of trajectories to an adjacent or neighboring
chaotic region, as in the setting of Refs. \cite{grigolini1}, \cite
{grigolini2} or as in trajectories in conservative maps with weakly
developed chaotic regions \cite{baldovin4}. Hence there is no reappearance
of trajectories from chaotic regions that would cause the relaxation from
the nonextensive regime with vanishing ordinary Lyapunov exponent to an
extensive regime with a positive one at some crossover iteration time $\tau $%
.

The links we have exhibited through exact analytical results between the
various properties of one-dimensional non-linear maps at critical conditions
provide a clear-cut corroboration, with a universal attribute, in the RG
sense, of the validity of non-extensive statistics at such states.

Acknowledgments. I would like to thank C. Tsallis and F. Baldovin for useful
discussions and comments. I gratefully acknowledge the hospitality of the
Centro Brasileiro de Pesquisas Fisicas where this work was carried out and
the financial support given by the CNPq processo 300894/01-5 (Brazil). This
work was also partially supported by CONACyT grant P-40530-F (Mexico).

\end{document}